# Form Factors for $B \to \pi l \bar{\nu}_l$ and $B \to K^* \gamma$ Decays on the Lattice

*UKQCD Collaboration*

**D.R. Burford, H.D. Duong, J.M. Flynn, J. Nieves**

Department of Physics, University of Southampton, Southampton SO17 1BJ, UK

**B.J. Gough**

Theoretical Physics MS106, Fermilab, Batavia IL 60510, USA

**N. M. Hazel**

Department of Physics & Astronomy, The University of Edinburgh, Edinburgh EH9 3JZ, Scotland

**H.P. Shanahan**

Department of Physics & Astronomy, University of Glasgow, Glasgow G12 8QQ, Scotland

## Abstract

We present a unified method for analysing form factors in $B \to \pi l \bar{\nu}_l$ and $B \to K^* \gamma$ decays. The analysis provides consistency checks on the $q^2$ and $1/M$ extrapolations necessary to obtain the physical decay rates. For the first time the $q^2$ dependence of the form factors is obtained at the $B$ scale. In the $B \to \pi l \bar{\nu}_l$ case, we show that pole fits to $f^+$ may not be consistent with the $q^2$ behaviour of $f^0$, leading to a possible factor of two uncertainty in the decay rate and hence in the value of $|V_{ub}|^2$ deduced from it. For $B \to K^* \gamma$, from the combined analysis of form factors $T_1$ and $T_2$, we find the hadronisation ratio $R_{K^*}$ of the exclusive $B \to K^* \gamma$ to the inclusive $b \to s \gamma$ rates is of order 35% or 15% for constant and pole-type behaviour of $T_2$ respectively.

# 1 Introduction

In this paper we describe a method of fitting lattice results for the matrix elements of the "heavy to light" decays $B \to \pi l \bar{\nu}_l$ and $B \to K^* \gamma$ to extract form factors. For $B \to \pi l \bar{\nu}_l$, we need the form factor as a function of $q^2$, where $q$ is the four-momentum transferred to the leptons. The $B \to \pi l \bar{\nu}_l$ decay rate can then be predicted, allowing information on the Cabibbo–Kobayashi–Maskawa (CKM) matrix element $V_{ub}$ to be extracted once an experimental measurement is made. For $B \to K^* \gamma$ we need a result at the on-shell point, $q^2 = 0$, where $q$ is the photon four-momentum. This decay has recently been measured by CLEO [1], and is a place where physics beyond the standard model, in particular supersymmetry, may be tested.

In both cases, we have lattice calculations performed with heavy quarks at around the charm mass which need to be extrapolated to the bottom mass [2, 3, 4, 5, 6, 7, 8, 9]. After this extrapolation we typically obtain the form factors close to the maximum squared momentum transfer, $q^2_{\max}$. This leaves a large extrapolation to be performed to reach low values of $q^2$ which dominate the phase space integral for $B \to \pi l \bar{\nu}_l$, or $q^2 = 0$ which is needed to find the rate for $B \to K^* \gamma$.

Our method is to use the lattice data as fully as possible. We will show how heavy quark symmetry makes it possible to obtain information about the $q^2$ dependence of the form factors for $B$ decays. The $q^2$ behaviour can now be tested, independently of the heavy quark extrapolation. Furthermore, we will use kinematic constraints to guide our extractions of the form factors. Consequently, in our final results, we have better control over the extrapolations and can check their self-consistency.

# 2 Form Factors

Here we give the standard expressions defining form factors for the two decays.

## 2.1 $B \to \pi l \bar{\nu}_l$

For this decay we need the vector current matrix element between $B$ and $\pi$ states:

$$\langle \pi(k) | \overline{u} \gamma_\mu b | B(p) \rangle = \left( p + k - q \frac{m_B^2 - m_\pi^2}{q^2} \right)_\mu f^+(q^2) + q_\mu \frac{m_B^2 - m_\pi^2}{q^2} f^0(q^2). \quad (1)$$

where $q = p - k$. In the rest frame of the decay products, $f^+$ and $f^0$ correspond to $1^-$ and $0^+$ exchanges respectively. At $q^2 = 0$ we have the constraint that

$$f^+(q^2{=}0) = f^0(q^2{=}0), \quad (2)$$

since the matrix element in equation (1) is non-singular at this kinematic point.

For zero-recoil, where the $B$ and $\pi$ four-velocities are equal, and $q^2 = q^2_{\max} = (m_B - m_\pi)^2 = 26.4 \,\text{GeV}^2$, only the time component of the matrix element is nonvanishing. At this point, the coefficient of $f^+$ vanishes and $f^0(q^2_{\max})$ alone can be determined.

The decay rate is dominated by $f^+$ since the contribution of $f^0$ vanishes for massless leptons.

## 2.2 $B \to K^*\gamma$

The matrix element of interest is

$$\langle K^*(k,\eta)|\bar{s}\sigma_{\mu\nu}q^\nu b_R|B(p)\rangle = \sum_{i=1}^{3} C_\mu^i T_i(q^2), \quad (3)$$

where $q = p - k$ as above, $\eta$ is the $K^*$ polarisation vector and

$$C_\mu^1 = 2\epsilon_{\mu\nu\lambda\rho}\eta^\nu p^\lambda k^\rho, \quad (4)$$

$$C_\mu^2 = \eta_\mu(m_B^2 - m_{K^*}^2) - \eta \cdot q(p+k)_\mu, \quad (5)$$

$$C_\mu^3 = \eta \cdot q \left(q_\mu - \frac{q^2}{m_B^2 - m_{K^*}^2}(p+k)_\mu\right). \quad (6)$$

For an on-shell photon with $q^2 = 0$, $T_3$ does not contribute to the $B \to K^*\gamma$ amplitude and $T_1$ and $T_2$ are related by,

$$T_1(q^2{=}0) = iT_2(q^2{=}0). \quad (7)$$

Hence, for $B \to K^*\gamma$, we need to determine $T_1$ and/or $T_2$ at the on-shell point.

The zero-recoil point occurs when $q^2 = q^2_{\max} = (m_B - m_{K^*})^2 = 19.2\,\text{GeV}^2$. At this point the contributions from $T_1$ and $T_3$ vanish, so $T_2(q^2_{\max})$ alone can be determined. The expected $t$-channel exchange particle for the $T_2$ form factor is the $1^+$ $B_{s1}$ state.

## 3 Heavy Quark Symmetry

We now turn to the predictions of heavy quark effective theory (HQET) for these matrix elements in the limit of infinite $b$-quark mass [10]. It is convenient to use the four velocities, $v$ and $v'$ of the mesons, defined by

$$p = Mv \quad \text{and} \quad k = mv' \quad (8)$$

where $M$ is the mass of the initial pseudoscalar ($B$) and $m$ is the mass of the final vector ($K^*$) or pseudoscalar ($\pi$) meson. We also use the variable

$$\omega = v \cdot v' = \frac{M^2 + m^2 - q^2}{2Mm} \quad (9)$$

so that zero-recoil occurs at $\omega = 1$.

The zeroth order HQET predictions for the $B \to \pi l\bar{\nu}_l$ form factors are:

$$f^+(q^2, x) = \frac{\sqrt{x}}{2}\left(\theta_1^P(\omega) - \frac{\theta_2^P(\omega)}{x}\right), \quad (10)$$

$$f^0(q^2, x) = \frac{\sqrt{x}}{1-x^2}\left((1-x\omega)\theta_1^P(\omega) - (\omega - x)\theta_2^P(\omega)\right). \quad (11)$$

where $x = m/M$. The functions $\theta_{1,2}^P$ depend on the light degrees of freedom, but do not depend on the heavy quark mass. They are also independent of the gamma-matrix structure of the current.

For $B \to K^*\gamma$ the corresponding form factor predictions are [11]:

$$T_1(q^2, x) = \frac{\sqrt{x}}{2}\left(\theta_1^V(\omega) - \frac{\theta_2^V(\omega)}{x}\right), \qquad (12)$$

$$iT_2(q^2, x) = \frac{\sqrt{x}}{1-x^2}\left((1-x\omega)\theta_1^V(\omega) - (\omega - x)\theta_2^V(\omega)\right), \qquad (13)$$

The expressions are identical in form to those for $B \to \pi l \bar{\nu}_l$, but the functions $\theta_{1,2}^V$ are different since they depend on different light degrees of freedom. We note in passing that the same two $\theta_{1,2}^V$ functions govern the HQET prediction for the $V$ and $A_1$ form factors in heavy pseudoscalar to light vector semileptonic decays proceeding via the left-handed vector current.

These predictions will be subject to perturbatively calculable renormalisation by strong interactions and corrections from terms with higher powers of $1/M$. However, they provide us with a starting point for our fitting procedure.

For a fixed value of $\omega$, the predictions in equations (10), (11), (12) and (13), give simple scaling laws in the heavy quark limit, $M \to \infty$ ($x \to 0$). In particular, at $q^2_{\max}$ we obtain

$$\begin{array}{lll} f^+(q^2_{\max}) \sim M^{1/2} & & T_1(q^2_{\max}) \sim M^{1/2} \\ f^0(q^2_{\max}) \sim M^{-1/2} & \text{and} & T_2(q^2_{\max}) \sim M^{-1/2} \end{array} \qquad (14)$$

The point with $q^2 = 0$ does *not* correspond to fixed $\omega$ as $M \to \infty$. This means we have to guess the $q^2$ dependence of the form factors if we are to extract scaling behaviour at $q^2 = 0$. Pole dominance ideas suggest that,

$$f(q^2) = \frac{f(0)}{(1 - q^2/M_f^2)^{n_f}} \qquad (15)$$

for $f = f^+, f^0, T_1, T_2$, where $M_f$ is a mass that is equal to $M$ plus $1/M$ corrections and $n_f$ is a power. Since $1 - q^2_{\max}/M_f^2 \sim 1/M$ for large $M$, the combination of heavy quark symmetry and the form factor relations at $q^2 = 0$ implies that $n_{f^+} = n_{f^0} + 1$ and $n_{T_1} = n_{T_2} + 1$. So, if we fit $f^+$ or $T_1$ to single pole forms, then $f^0$ or $T_2$ should be constant in $q^2$. To have a single pole form for $f^0$ or $T_2$ necessitates double pole (or "dipole") forms for $f^+$ or $T_1$. These two types of behaviour correspond to

$$f^+(0) \text{ or } T_1(0) \sim \begin{cases} M^{-1/2} & \text{single pole} \\ M^{-3/2} & \text{double pole} \end{cases}. \qquad (16)$$

In passing from pole/constant to dipole/pole behaviour we are allowing $f^0$ for $B \to \pi l \bar{\nu}_l$ or $T_2$ for $B \to K^*\gamma$ to acquire curvature in $q^2$. As we will see, both $f^0$ and $T_2$ have weak dependence on $q^2$ for the measured points: in this case constrained multipole fits, for example tripole/dipole, to these form factors are nearly indistinguishable in the region of $q^2$ between 0 and $q^2_{\max}$, since one can compensate higher powers by changes in the fitted mass parameter. For this reason, we will show results for pole/constant and dipole/pole fits only.

## 4 Fitting Lattice Data

From lattice calculations with propagating (rather than static) quarks, we can obtain matrix elements for heavy quarks around the charm mass over a range of $q^2$ straddling $q^2 = 0$. In extracting form factors from these matrix elements, we can reach $q^2_{\max}$ for $f^0$ and $T_2$ only. We need to extrapolate in the heavy mass $M$ to scales around the $b$ quark mass. Such scaling in $M$ is simple for fixed $\omega$, but, at the $B$ scale, produces a range of $q^2$ values near $q^2_{\max}$ and far from $q^2 = 0$. We are therefore faced with a large extrapolation to $q^2 = 0$.

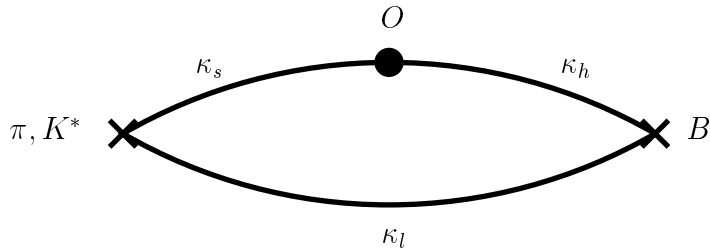

Figure 1: Labelling of quark hopping parameters for three point correlator calculation.

The results described below come from 60 $SU(3)$ gauge configurations generated by the UKQCD collaboration on a $24^3 \times 48$ lattice at $\beta = 6.2$ in the quenched approximation. The $\mathcal{O}(a)$ improved Sheikholeslami–Wohlert [12] action was used for fermions, with "rotated" fermion fields appearing in all operators used for correlation function calculations [13]. The inverse lattice spacing determined from the string tension is $a^{-1} = 2.73(5)$ GeV [14].

Three point correlators of the heavy-to-light two fermion operators with a heavy pseudoscalar meson (the "$B$" meson) and a light pseudoscalar or vector meson were calculated, as illustrated in figure 1. Matrix elements were extracted from these correlators by the method detailed in [9, 15, 16, 17]. Four heavy quark hopping parameters, $\kappa_h = 0.121, 0.125, 0.129, 0.133$, were used. For the propagator connecting the current operator to the light meson operator, two kappa values, $\kappa_s = 0.14144, 0.14226$, were available. The subscript $s$ is for strange: these kappa values straddle that for the strange quark, $0.1419(1)$ [18]. For $\kappa_h = 0.121, 0.129$, we used three light "spectator" hopping parameters, $\kappa_l = 0.14144, 0.14226, 0.14262$, and for $\kappa_h = 0.125, 0.133$ we used $\kappa_l = 0.14144$ only. The critical hopping parameter at this $\beta$ is $\kappa_{\mathrm{crit}} = 0.14315(1)$ [18].

The lattice calculations were performed with the heavy meson spatial momentum of magnitude 0 or 1, in lattice units of $\pi/12a$. The momentum injected at the operator insertion was varied to allow the modulus of the light meson spatial momentum to take values up to $\sqrt{3}$ in lattice units (although some of the momentum choices were too noisy to be used in fits). We refer to each combination of light and heavy meson three-momenta as a channel with the notation $|\mathbf{p}| \to |\mathbf{k}|$ in lattice units (for example $0 \to 1$ or $1 \to 1_\perp$ where the subscript $\perp$ indicates that $\mathbf{p}$ and $\mathbf{k}$ are perpendicular).

The results below have been obtained using uncorrelated fits for the extrapolations in the heavy quark mass and in $q^2$. The extraction of the form factors from the three-point correlation function data used correlated fits [9, 15]. Statistical errors are 68% confidence limits obtained from 1000 bootstrap samples.

To make the best use of HQET, we pick momentum combinations which keep $\omega$ constant or nearly constant as the heavy mass varies. This will allow us to scale linearly or quadratically in $1/M$ from the charm to the bottom mass scale and gives us the form factors for the $B$ decays as a function of $\omega$. Now we convert $\omega$ to $q^2$ and fit to assumed forms for the $q^2$ dependence at the $B$ scale, consistent with the relations $f^+(0) = f^0(0)$ and $T_1(0) = iT_2(0)$.

| momenta | | $\kappa_h$ | | | | |
|---|---|---|---|---|---|---|
| $|\mathbf{p}|$ | $|\mathbf{k}|$ | 0.121 | 0.125 | 0.129 | 0.133 | $\omega_{\text{ave}}$ |
| 0 | 0 | 1.00 | 1.00 | 1.00 | 1.00 | 1.00 |
| 0 | 1 | 1.33 | 1.33 | 1.33 | 1.33 | 1.33 |
| 0 | $\sqrt{2}$ | 1.60 | 1.60 | 1.60 | 1.60 | 1.60 |
| 1 | 0 | 1.04 | 1.05 | 1.06 | 1.09 | 1.06 |
| 1 | $1_\perp$ | 1.38 | 1.40 | 1.42 | 1.45 | 1.41 |
| 1 | $\sqrt{2}_\perp$ | 1.66 | 1.68 | 1.70 | 1.74 | 1.69 |

Table 1: Values of $\omega$ and their averages for $B \to \pi l \bar{\nu}_l$ with various $\kappa_h$ values and $\kappa_s = \kappa_l = 0.14144$. We average over values of the light meson momentum $\mathbf{k}$ when possible, but for the heavy meson only two momenta are available ($\mathbf{p} = (0,0,0)$ and $(1,0,0)$ in lattice units of $\pi/12a$).

| momenta | | $\kappa_h$ | | | | |
|---|---|---|---|---|---|---|
| $|\mathbf{p}|$ | $|\mathbf{k}|$ | 0.121 | 0.125 | 0.129 | 0.133 | $\omega_{\text{ave}}$ |
| 0 | 0 | 1.00 | 1.00 | 1.00 | 1.00 | 1.00 |
| 0 | 1 | 1.22 | 1.22 | 1.22 | 1.22 | 1.22 |
| 1 | 0 | 1.04 | 1.05 | 1.07 | 1.09 | 1.06 |
| 1 | $1_\perp$ | 1.27 | 1.28 | 1.30 | 1.33 | 1.30 |

Table 2: Values of $\omega$ for $B \to K^*\gamma$ for various momentum channels and $\kappa_h$ values, with their average for each channel. Calculated with $m = m_{K^*}$ for $\kappa_s = 0.1419$ and $\kappa_l = 0.14144$.

Since
$$\omega = v \cdot v' = \frac{E_M E_m - \mathbf{p} \cdot \mathbf{k}}{Mm}, \tag{17}$$
we can select channels where $\mathbf{p} \cdot \mathbf{k} = 0$, so that
$$\omega = \frac{E_m}{m}\left(1 + \frac{\mathbf{p}^2}{2M^2} + \cdots\right). \tag{18}$$

We see that $\omega$ is independent of $M$ when the heavy meson is at rest and the light meson momentum is fixed. When $|\mathbf{p}| = 1$ in lattice units, the change in $\omega$ is only about 6% in a given channel for our range of heavy kappa values. Given that this error is comparable with others from discretisation effects, quenching and so on, this should not preclude inclusion of these channels in fitting. We take the actual $\omega$ as an average of the $\omega$'s for the four heavy kappas. In tables 1 and 2 we show the channels used and the corresponding $\omega$ values.

## 4.1 $B \to \pi l \bar{\nu}_l$ Results

For this exploratory study, not enough spectator quark kappa values were available for all heavy quark kappa values to allow reliable chiral extrapolations. Therefore, our aim in this section is not to quote realistic values for the form factors governing the $B \to \pi l \bar{\nu}_l$ decay, but to illustrate the main features of the analysis method proposed in this paper. All results presented in this section are for the case $\kappa_s = \kappa_l = 0.14144$.

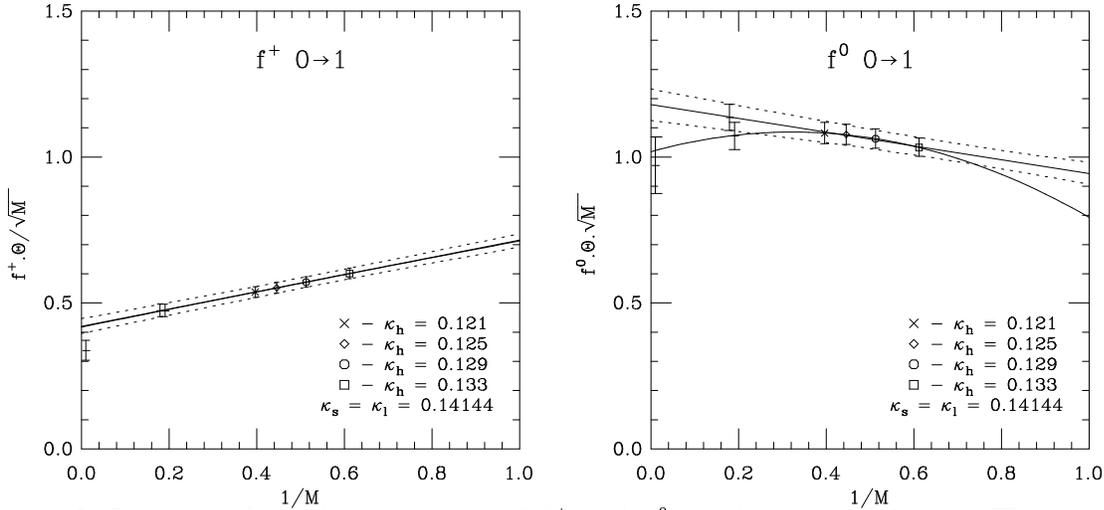

Figure 2: Linear and quadratic scaling of $f^+$ and $f^0$ for the channel $0 \to 1$. The perturbative value, 0.83 [19], was used for the vector current renormalisation constant, $Z_V$. $M$ is given in units of GeV. The dotted lines show the statistical error on the linear fits.

We first scale the form factors measured at fixed $\omega$ to the $B$ mass, using both linear and quadratic dependence on $1/M$ to extrapolate:

$$f^+ \Theta/\sqrt{M} = \begin{cases} \gamma_+ \left(1 + \frac{\delta_+}{M}\right) & \text{linear} \\ \gamma_+ \left(1 + \frac{\delta_+}{M} + \frac{\epsilon_+}{M^2}\right) & \text{quadratic} \end{cases} \quad (19)$$

$$f^0 \Theta \sqrt{M} = \begin{cases} \gamma_0 \left(1 + \frac{\delta_0}{M}\right) & \text{linear} \\ \gamma_0 \left(1 + \frac{\delta_0}{M} + \frac{\epsilon_0}{M^2}\right) & \text{quadratic} \end{cases} \quad (20)$$

where $\Theta$ comes from the leading logarithmic factors and is chosen to be 1 at the $B$ mass,

$$\Theta = \Theta(M/m_B) = \left(\frac{\alpha_s(M)}{\alpha_s(m_B)}\right)^{\frac{2}{\beta_0}}. \quad (21)$$

with $\beta_0 = 11$ in the quenched approximation and $\Lambda_{QCD} = 200$ MeV. An example of the scaling is shown in figure 2 for the channel $0 \to 1$. We quote results from the linear extrapolation, but, as an indication of the systematic error, incorporate the difference between linear and quadratic extrapolations in quadrature with the statistical error. We have not incorporated any systematic error due to ambiguities in the determination of $a^{-1}$. We note in passing that taking the chiral limit would also increase our errors.

In figure 2 we show also the form factors in the static limit obtained from [20]. The discrepancy between the extrapolation and the static point could be due either to discretisation errors for the propagating quarks or to excited state contributions in the static results. The limited agreement gives some confidence in our extrapolations.

The extrapolated form factors were fitted separately to different $q^2$ dependences: $f^+$ to a dipole and a pole, $f^0$ to a pole and a constant. Momentum channels $0 \to 1$, $0 \to \sqrt{2}$, $1 \to 0$, $1 \to 1_\perp$, $1 \to \sqrt{2}_\perp$ and, for $f^0$ only, $0 \to 0$, were used. The fits are illustrated in figure 3. The dipole/pole combination for $f^+$, $f^0$ is favoured over the pole/constant or indeed over fitting both form factors to single poles, because this combination comes closest to obeying the relation $f^+(0) = f^0(0)$.

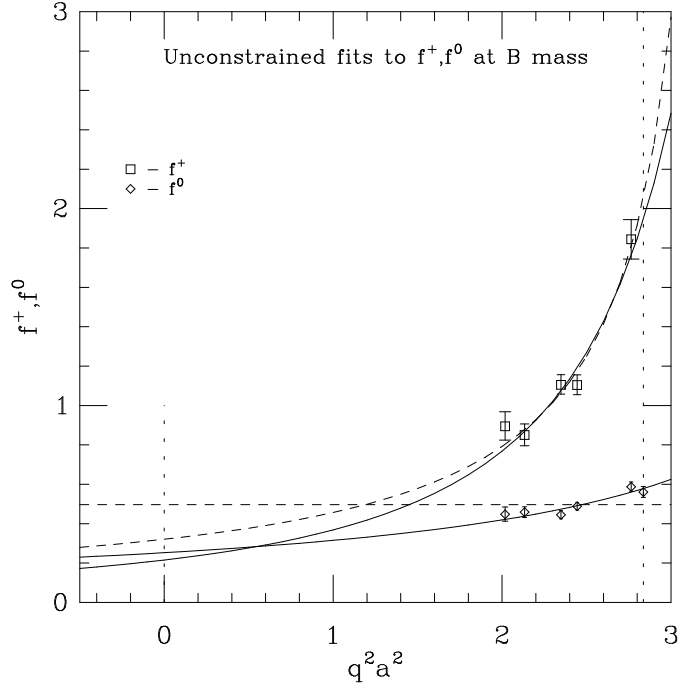

Figure 3: Unconstrained dipole/pole fits (solid lines) and pole/constant fits (dashed lines). The dotted lines mark $q^2 = 0$ and $q^2_{\max}$. The $\chi^2$/dof for the fits were comparable save for the constant fit to $f^0$ which was about 4 times greater.

Dipole/pole and pole/constant fits constrained to satisfy $f^+(0) = f^0(0)$ are shown in figure 4 with numerical values given in table 3. Again the dipole/pole fit is preferred over the pole/constant fit.

Constrained fits of $f^{+,0}(q^2)$ were also carried out for each of the four $\kappa_h$'s using momentum channels $1 \to -1$, $1 \to 1$ and $1 \to (1,1,0)$ in addition to those used above. The resultant interpolations to $f(q^2 = 0)$ (we drop the superscripts on $f^+$ and $f^0$ when they are constrained to agree at $q^2 = 0$) were scaled appropriately to the $B$ mass according to equation (16). These are the "burst points" shown in figure 4, with values given in table 3. The agreement between the two methods shows the self consistency of using dipole/pole or pole/constant constrained fits.

In previous lattice calculations [2, 3] the $f^+$ form factor was determined using the $0 \to 1$ momentum channel and assuming a $q^2$ dependence given by the exchange of the $B^*$ resonance in the $t$-channel. We will refer to this as the "pole procedure" below. Here we use the physical $B_s^*$ mass (5.46 GeV) for the pole since this corresponds roughly to the quark kappa values employed. This procedure gives $f^+(0) = 0.43 \pm 0.02$, in reasonable agreement with the constrained pole/constant fit.

The results of table 3 and figure 4 suggest that a constrained dipole/pole fit is more appropriate than a pole/constant fit. The result of using such a fit for $f^+$ is to reduce the value for the decay rate from that using a pole behaviour for $f^+$, as shown in table 4. As a consequence, the extracted value of the quark mixing parameter $|V_{ub}|^2$ changes by a factor of two.

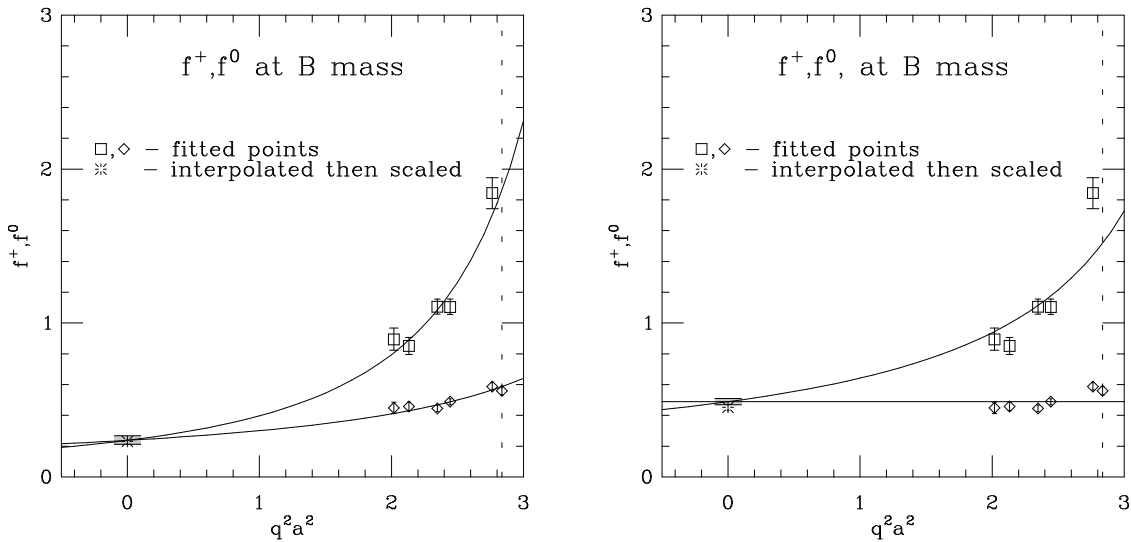

Figure 4: Constrained dipole/pole (left) and pole/constant (right) fits to $f^+$ and $f^0$ at the $B$ mass. The burst point, is from $f(0)$'s obtained from constrained fits at the four $\kappa_h$'s scaled appropriately to the $B$ mass. The shaded bands mark the error on $f(0)$ obtained from the fits. The dotted line marks $q^2_{\max}$.

## 4.2  $B \to K^* \gamma$ Results

The lack of spectator quark kappa values is less of a problem here than for $B \to \pi l \bar{\nu}_l$. It has been shown that the form factors' dependence on the spectator quark mass is mild [9]. As in reference [9] we interpolate $\kappa_s$ to the physical strange kappa value, 0.1419(1), but neglect any dependence of the form factors on the light spectator quark mass. Therefore we present results only for $\kappa_l = 0.14144$. We then scale to the $B$ mass using a similar procedure to that explained for $B \to \pi l \bar{\nu}_l$.

For $T_1$ we use the following momentum channels: $0 \to 1$, $1 \to 0$ and $1 \to 1_\perp$. For $T_2$ we can use the $0 \to 0$ channel in addition to these. We do not use the channels $0 \to \sqrt{2}$ and $1 \to \sqrt{2}_\perp$ as both the statistical and systematic errors on the points are large. We fit $T_1$ and $T_2$ simultaneously, imposing $T_1(0) = iT_2(0)$, with pole/constant or dipole/pole forms. These fits are shown in figure 5 with numerical values given in table 5.

| fit type | burst pt. | $f(0)$ | $m_{f^+}$ | $m_{f^0}$ | $\chi^2/\mathrm{dof}$ |
|---:|---:|---:|---:|---:|---:|
| dipole/pole | $0.23 \pm 0.02$ | $0.24^{+0.04}_{-0.03}$ | $5.7 \pm 0.1$ | $6.0^{+0.3}_{-0.2}$ | 1.5 |
| pole/const | $0.45 \pm 0.02$ | $0.49 \pm 0.02$ | $5.6 \pm 0.3$ | — | 6.1 |

Table 3: Constrained dipole/pole vs pole/constant fits at the $B$ mass for $B \to \pi l \bar{\nu}_l$ with $\kappa_s = \kappa_l = 0.14144$. The burst point, see figure 4, is from $f(0)$'s obtained from constrained fits at the four $\kappa_h$'s scaled appropriately to the $B$ mass. Fitted masses are in GeV.

| fit type | $\Gamma(B \to \pi l \bar{\nu}_l)/|V_{ub}|^2 10^{12}\ \mathrm{s}^{-1}$ |
|---|---|
| dipole/pole | $5.3^{+1.2}_{-1.0}$ |
| pole/const | $13.0 \pm 1.6$ |
| pole procedure | $10.2 \pm 3.6$ |

Table 4: Decay rates calculated using different forms for $f^+(q^2)$. The result in the last row has been obtained using the $0 \to 1$ momentum channel and a single pole form for $f^+$, as explained in the text.

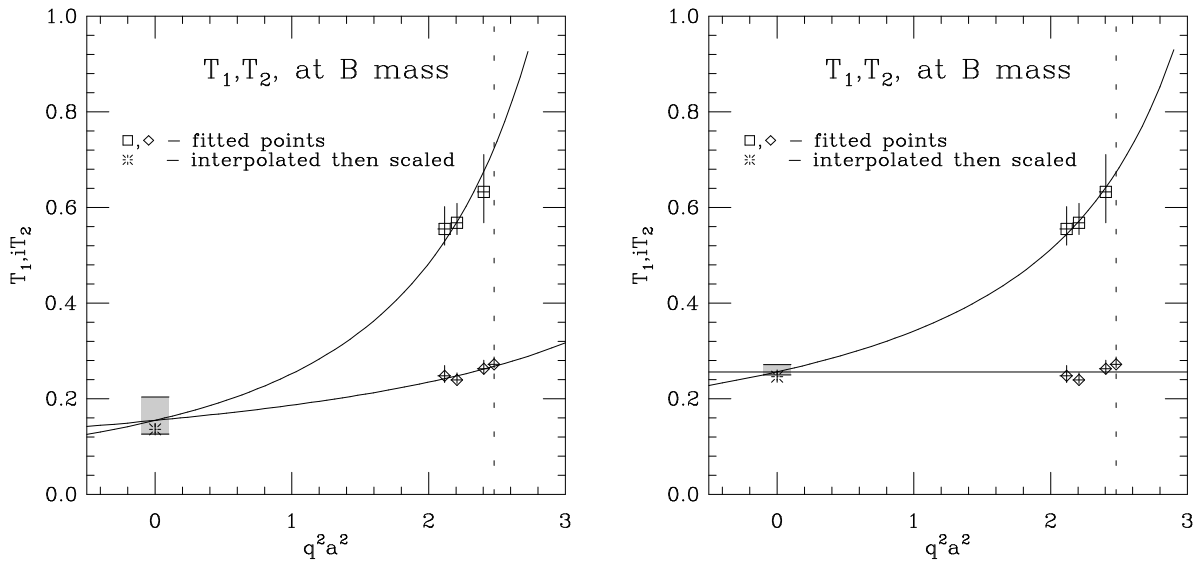

Figure 5: Constrained dipole/pole (left) and pole/constant (right) fits for $T_1$ and $T_2$ at the $B$ mass. The burst point, is from $T(0)$'s obtained from constrained fits at the four $\kappa_h$'s scaled appropriately to the $B$ mass. The shaded bands mark the errors on $T(0)$ obtained from the fits. The dotted line marks $q^2_{\max}$.

Alternatively, for each $\kappa_h$, we fit $T_1$ and $T_2$ as functions of $q^2$, with $T_1(0) = iT_2(0)$ imposed, adding channels $1 \to -1$ and $1 \to 1$ to the above list. This gives $T(q^2=0)$ (we let $T(0)$ denote $T_1(0) = iT_2(0)$). These points are extrapolated to the $B$ scale according to equation (16). The $q^2 = 0$ point is the burst point plotted in figure 5 and referred to in table 5.

As can be seen in figure 5 and table 5, fitting form factors using the two methods described above produces consistent results. The burst points should agree with the results of reference [9]. For dipole/pole fits, this is indeed the case, but for pole/constant fits a much lower value was found for $T(0)$ in [9]. We believe this difference arises from our imposition of the constraint $T_1(0) = iT_2(0)$, which has reduced the curvature in the values of $T(0)$ as a function of $1/M$. In [9] the curvature produced a low value of the form factor at $q^2 = 0$ at the $B$ scale, which is inconsistent with the value of $T_2(q^2_{\max})$ if $T_2$ is constant.

We have checked that higher power pole fit combinations, for example tripole/dipole,

| fit type | burst pt. | $T(0)$ | $m_{T_1}$ | $m_{T_2}$ | $\chi^2/\text{dof}$ | $R_{K^*}$ |
|---|---|---|---|---|---|---|
| dipole/pole | $0.14 \pm 0.03$ | $0.15^{+0.07}_{-0.06}$ | $5.8^{+0.6}_{-0.5}$ | $7^{+3}_{-2}$ | 0.4 | $(13^{+14}_{-10})\%$ |
| pole/const | $0.25 \pm 0.02$ | $0.26^{+0.02}_{-0.01}$ | $5.4 \pm 0.1$ | — | 0.9 | $(35^{+4}_{-2})\%$ |

Table 5: Dipole/pole vs pole/constant fits for $B \to K^*\gamma$ for $\kappa_s = \kappa_l = 0.14144$. The burst point, see figure 5, is from $T(0)$'s obtained from constrained fits at the four $\kappa_h$'s scaled appropriately to the $B$ mass. Fitted masses are given in GeV. $R_{K^*}$ is the hadronisation ratio defined in the text, where a value of $m_b = 4.65\,\text{GeV}$ [21] is used.

give results consistent with the dipole/pole fit.

In contrast to the $B \to \pi l \bar{\nu}_l$ case, our final results do not allow us to favour any particular fit combination. Hence, from this study, we can conclude only that $T(0)$ lies in the range 0.09 to 0.28. This situation is common to current lattice calculations of $B \to K^*\gamma$ form factors [6, 7, 9]. With higher statistics and smaller lattice spacings (allowing larger spatial momenta and hence smaller $q^2$ values), the method proposed here should be able to differentiate between $q^2$ behaviours and eliminate the present uncertainty.

For comparison with experiment it is useful to consider the hadronisation ratio, given up to $\mathcal{O}(1/m_b^2)$ corrrections by [21]

$$R_{K^*} = \frac{\Gamma(B \to K^*\gamma)}{\Gamma(b \to s\gamma)} \qquad (22)$$

$$= 4\left(\frac{m_B}{m_b}\right)^3 \left(1 - \frac{m_{K^*}^2}{m_B^2}\right)^3 |T(0)|^2, \qquad (23)$$

in which many of the theoretical uncertainties in relating the form factor to the branching ratio cancel.[1] We report values for $R_{K^*}$ in table 5. The experimental value is $R_{K^*} = (19 \pm 13)\%$ [1, 23]. It appears that the dipole/pole fit is in better agreement with the hadronisation ratio data, but the errors in both the fit and the experimental measurement are large.

## 5 Conclusions

We have presented a unified method for analysing form factors in $B \to \pi l \bar{\nu}_l$ and $B \to K^*\gamma$ decays. We have extrapolated as many momentum channels as possible at nearly fixed $\omega$ to give the maximum information on the $q^2$ dependence of the form factors at the $B$ scale. For the first time, we have been able to fit this $q^2$ dependence directly. This analysis, combined with the procedure of extrapolating to $q^2 = 0$ and then scaling to the $B$ mass, provides consistency checks on the extrapolations necessary to obtain the physical decay rates. The imposition of the constraint at $q^2 = 0$ in all our fits is novel and provides a framework for the best utilisation of the lattice data.

Our main conclusions are that for the $B \to \pi l \bar{\nu}_l$ case, pole fits to $f^+$ may not be consistent with the $q^2$ behaviour of $f^0$, leading to a possible factor of 2 uncertainty in future determinations of $|V_{ub}|^2$ from this decay. For $B \to K^*\gamma$, with the current data, no particular $q^2$ behaviour at the $B$ scale is favoured. The resulting uncertainty in $T(0)$ is large, limiting the usefulness of measurements of this decay for constraining the standard model or new physics.

---
[1] The theoretical prediction for this ratio may be subject to long-distance effects [22].

The problems associated with the $q^2$ extrapolation will be generic to the study of any decay of a $B$ meson into a light meson plus leptons or a photon. Direct simulations of the $B$ meson on the lattice, using NRQCD or the heavy Wilson action (as proposed by the Fermilab group), will face the same problems, if they are limited to a small range of $q^2$ around $q^2_{\max}$.

These considerations do not modify previous results for charmed meson semileptonic decays, because the $q^2$ range available from lattice simulations covers the physically accessible range. Pole and dipole fits for $f^+$ agree over the physical range of $q^2$ for charm decays.

The analysis method presented here, together with measurements closer to $q^2 = 0$, will provide an unambiguous determination of the $q^2$ behaviour of the form factors. In this study, we could not use the $0, 1 \to \sqrt{3}$ (and $0, 1 \to \sqrt{2}$ for $B \to K^*\gamma$) channels which were already too noisy. Future lattice simulations with higher statistics and smaller lattice spacings at fixed physical volume, will allow more momentum channels to be used, thereby increasing the usable range of $q^2$.

## Acknowledgments

We thank Guido Martinelli for emphasising the need for consistency in fitting form factors for $B \to K^*\gamma$ and for alerting us to reference [22]. Rajan Gupta and Tanmoy Bhattacharya [24] have independently noted the similarities between $B \to \pi l \bar{\nu}_l$ and $B \to K^*\gamma$ form factor extractions. We also thank Chris Sachrajda, Hartmut Wittig and other members of the UKQCD collaboration for useful discussions. JN acknowledges the European Union for for their support through the award of a Postdoctoral Fellowship, contract No. CHBICT920066. We acknowledge the Particle Physics and Astronomy Research Council for travel support under grant GR/J98202.